\RequirePackage{lineno}
\documentclass[aps,prc,twocolumn, floatfix]{revtex4}
\usepackage{graphicx}
\usepackage{dcolumn}
\usepackage{bm}
\usepackage{mhchem}
\usepackage{color}
\usepackage{tcolorbox}
\usepackage{multirow}
\usepackage{float}

\begin{document}


\title{Violation of NCQ scaling in hadron elliptic flow in Au+Au collisions at $\sqrt{s_{NN}}=3.0-7.7  \,\mathrm{GeV}$
}

\author{Xun Zhu$^{1,2}$}
\author{Gao-Chan Yong$^{1,2,3}$}
\email[Corresponding author: ]{yonggaochan@impcas.ac.cn}

\affiliation{
$^1$Institute of Modern Physics, Chinese Academy of Sciences, Lanzhou 730000, China\\
$^2$School of Nuclear Science and Technology, University of Chinese Academy of Sciences, Beijing 100049, China\\
$^3$State Key Laboratory of Heavy Ion Science and Technology, Institute of Modern Physics, Chinese Academy of Sciences, Lanzhou 730000, China
}

\begin{abstract}
    We investigate the number-of-constituent-quark (NCQ) scaling of elliptic flow for various hadrons in non-central Au+Au collisions at \(\sqrt{s_{NN}} = 3.0\text{--}7.7\,\mathrm{GeV}\) using the AMPT model with string melting (SM) and pure hadron cascade (HC) modes. For the SM case, NCQ scaling is absent at \(\sqrt{s_{NN}} =3.0\,\mathrm{GeV}\) but is largely restored by \(\sqrt{s_{NN}} =3.9\,\mathrm{GeV}\). Although quark coalescence occurs at \(\sqrt{s_{NN}} =3.0\,\mathrm{GeV}\), the lack of NCQ scaling is attributed to the insufficient development of quark elliptic flow and the limited production of strange quarks and antiquarks. This finding suggests that NCQ scaling could not be considered a definitive signature of quark-gluon plasma (QGP) formation in the RHIC fixed-target energy region. For the HC case, as expected, no NCQ scaling is observed. However, a mass ordering in the elliptic flow emerges at \(\sqrt{s_{NN}} =4.5\,\mathrm{GeV}\), indicating that full thermalization may not be a prerequisite for mass ordering.
\end{abstract}

\maketitle

%
In relativistic heavy-ion collisions, a major research focus is the investigation of the Quantum Chromodynamics (QCD) phase transition and the properties of the quark-gluon plasma (QGP)~\cite{arslandok_HotQCD_2023, aidala_NewEra_2023}. However, due to the extremely short lifetime of the QGP, typically lasting only a few fm/c, and the extreme conditions present in such collisions, the QGP cannot be directly observed~\cite{fukushima_QCDMatter_2012, heinz_CollectiveFlow_2013b}. Consequently, identifying observables sensitive to the phase transition is crucial. Among these, elliptic flow ($v_2$) has emerged as a particularly powerful observable. In non-central collisions, the initial spatial overlap region forms an almond shape, leading to an anisotropic pressure gradient that results in the elliptic flow of final-state particles. Elliptic flow is quantified by the second-order Fourier coefficient of the particle azimuthal distribution and can be expressed as~\cite{poskanzer_MethodsAnalyzing_1998, voloshin_FlowStudy_1996}:
\begin{equation}
    v_2 = \left\langle \cos[2(\phi - \psi_r)] \right\rangle = \left\langle \frac{p_x^2 - p_y^2}{p_x^2 + p_y^2} \right\rangle,
\end{equation}
where $\phi$ is the particle's azimuthal angle, $\psi_r$ is the reaction plane angle, and $p_x$ and $p_y$ are the transverse momentum components along and perpendicular to the reaction plane, respectively. Elliptic flow develops during the early stages of the collision~\cite{sorge_EllipticalFlow_1997, ollitrault_AnisotropySignature_1992b, sorge_HighlySensitive_1999c} and is highly sensitive to the properties of the QGP~\cite{qin_TranslationCollision_2010, romatschke_ViscosityInformation_2007}, the nuclear equation of state (EoS)~\cite{lefevre_OriginElliptic_2018, snellings_EllipticFlow_2011, huovinen_QCDEquation_2010}, and the degree of thermalization within the system~\cite{bhalerao_EllipticFlow_2005b}. Thus, the study of elliptic flow is essential for advancing our understanding of the QCD phase transition and the nature of the QGP.

Experimental results from the Relativistic Heavy Ion Collider (RHIC)~\cite{adler_EllipticFlow_2002b} and the Large Hadron Collider (LHC)~\cite{abelev_EllipticFlow_2015a} reveal that differential elliptic flow exhibits distinct underlying mechanisms across different transverse momentum ($p_T$) regions. At low $p_T$, the elliptic flow of identified particles demonstrates a characteristic mass ordering~\cite{adler_AzimuthalAnisotropy_2002a, collaboration_IdentifiedParticle_2001a, adams_ParticleTypeDependence_2004a}, where heavier particles tend to have smaller $v_2$ values at a given $p_T$. In the relatively high $p_T$ region, however, elliptic flow shows a dependence on hadron type, with the magnitude of $v_2$ becoming proportional to the number of constituent quarks. This phenomenon is well described by quark coalescence or recombination models~\cite{jia_QuarkNumber_2007a, molnar_EllipticFlow_2003a, hwa_ScalingBehavior_2003}, which yield the following mathematical relationships:
\begin{equation}
    v_{2,M}(p_T) \approx 2 v_{2,q}\left(\frac{p_T}{2}\right), \quad
    v_{2,B}(p_T) \approx 3 v_{2,q}\left(\frac{p_T}{3}\right),
\end{equation}
where $v_{2,M}$ and $v_{2,B}$ represent the elliptic flow of mesons and baryons, respectively, and $v_{2,q}$ denotes the elliptic flow of constituent quarks. This scaling behavior is known as the number-of-constituent-quark (NCQ) scaling. The discovery of NCQ scaling suggests that collective flow develops at the partonic stage and provides strong evidence for quark coalescence as the dominant hadronization mechanism~\cite{jia_QuarkNumber_2007a}. As a result, the NCQ scaling of final-state particles in heavy-ion collisions is widely regarded as a key indicator of the formation of QGP during the collision process.

\begin{figure*}[tbh]
    \centering
    \includegraphics[width=15cm]{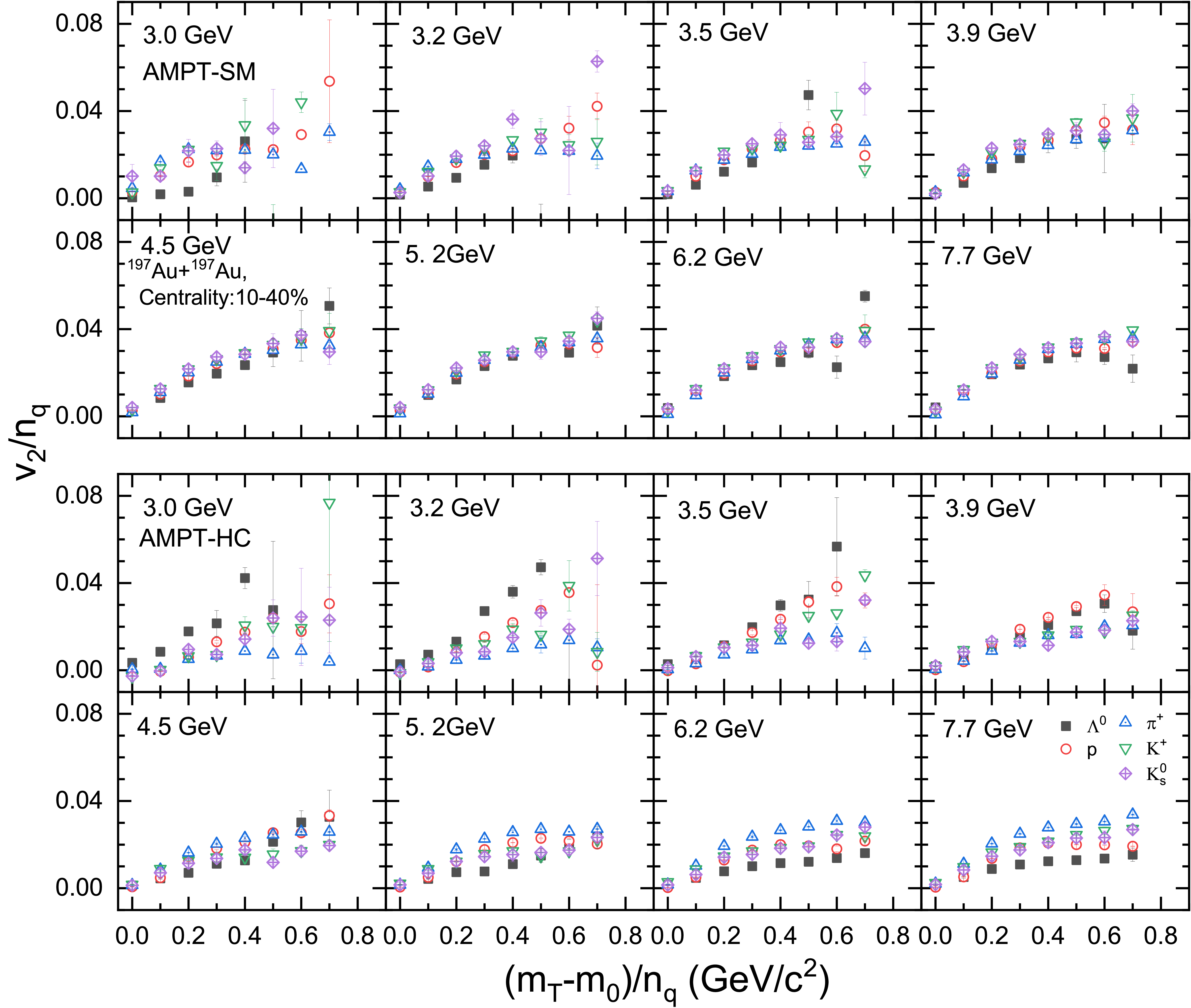}
    \caption{The $v_{2}/n_{q}$ of different hadrons as a function of $(m_{T}-m_{0})/n_{q}$ in Au+Au collisions at $\sqrt{s_{NN}}=3.0$--$7.7\,\mathrm{GeV}$ with a centrality of 10--40\%. Top: results from AMPT-SM. Bottom: results from AMPT-HC.} \label{1}
    \vspace{0.25cm}
\end{figure*}

The NCQ scaling was first observed in experiments at the top RHIC energy of $\sqrt{s_{NN}} = 200\,\mathrm{GeV}$~\cite{adams_ParticleTypeDependence_2004a, collaboration_ScalingProperties_2007, adare_DeviationQuarknumber_2012}. Over the subsequent two decades, this phenomenon was confirmed at energies up to several TeV at the LHC~\cite{abelev_EllipticFlow_2015} and down to $\sqrt{s_{NN}} = 7.7\,\mathrm{GeV}$ in the Beam Energy Scan (BES) program at RHIC~\cite{starcollaboration_EllipticFlow_2013}. Recent results from the BES-II fixed-target experiments show that NCQ scaling is observed at $\sqrt{s_{NN}} = 4.5\,\mathrm{GeV}$~\cite{abdallah_FlowInterferometry_2021} but is violated at $\sqrt{s_{NN}} = 3.0\,\mathrm{GeV}$~\cite{abdallah_DisappearancePartonic_2022}. Studies~\cite{abdallah_FlowInterferometry_2021, abdallah_DisappearancePartonic_2022, singh_InsightsPartonic_2025} suggest that this transition indicates a shift in the underlying dynamics of the system, from partonic interactions dominating at higher energies to hadronic interactions prevailing at lower energies. As the beam energy decreases, the lifetime of the partonic phase becomes shorter~\cite{sahoo_PROBINGQGP_2011}, leading to insufficient development of elliptic flow among patrons. This may affect the manifestation of the NCQ scaling. In addition, a significant increase in the baryon chemical potential is observed as the beam energy decreases~\cite{lan_AnisotropicFlow_2022}. Experimental data indicate that the elliptic flow difference between particles and their antiparticles increases linearly with the baryon chemical potential, and this difference is more pronounced for baryons than for mesons~\cite{starcollaboration_EllipticFlow_2013,starcollaboration_ObservationEnergyDependent_2013}. This phenomenon suggests a strong correlation between the violation of NCQ scaling and the baryon chemical potential~\cite{starcollaboration_EllipticFlow_2013}, which has been described by an extended coalescence model that incorporates the effect of baryon stopping \cite{dunlop_ConstituentQuark_2011}.

Current studies on the violation of NCQ scaling are primarily based on two phenomena observed in the early BES experiments~\cite{shi_EventAnisotropy_2013, adamczyk_MeasurementElliptic_2016, starcollaboration_EllipticFlow_2013}. First, the breaking of NCQ scaling between particles and antiparticles is observed at $\sqrt{s_{NN}} \leq 39\,\mathrm{GeV}$, which is commonly interpreted as an effect of the chiral magnetic effect~\cite{burnier_ChiralMagnetic_2011}, while some studies attribute it to baryon stopping \cite{dunlop_ConstituentQuark_2011}. Second, the elliptic flow of $\phi$ mesons is found to be significantly smaller than that of other mesons. This suppression has been interpreted as a consequence of incomplete thermalization of strange quarks~\cite{zhang_ProbePartonic_2012,dunlop_ConstituentQuark_2011} and/or the small hadronic scattering cross section of the $\phi$ meson, which limits its flow development during the hadronic interaction~\cite{adamczyk_ProbingParton_2016, mohanty_ProbeQCD_2009}. These two experimental phenomena suggest that the violation of NCQ scaling does not necessarily indicate the absence of a partonic phase. References \cite{dunlop_ConstituentQuark_2011, goudarzi_EvidenceCoalescence_2020} theoretically explained the above phenomenon. They consider the effect of baryon stopping, dividing quarks into produced and transported types, and suggests that quarks of different flavors exhibit different elliptic flows, leading to the conclusion that hadron elliptic flow cannot be simply scaled by the number of constituent quarks during coalescence. However, there is still a lack of systematic theoretical studies on the mechanism of NCQ scaling violation in the BES fixed-target energy region, particularly at $\sqrt{s_{NN}} = 3.0$--$4.5\,\mathrm{GeV}$. Therefore, we systematically investigate the applicability of NCQ scaling as a probe for phase transition signals in the energy range $\sqrt{s_{NN}} = 3.0$--$7.7\,\mathrm{GeV}$ and explore the key factors influencing its emergence. Because previous research has shown that the matter produced in heavy-ion collisions remains hadronic up to 3 GeV \cite{du23,wuzm23,abdallah_DisappearancePartonic_2022}; however, by 7.7 GeV, the matter produced is largely quark matter \cite{starcollaboration_EllipticFlow_2013}. Additionally, recent studies suggest that the hadron-quark phase transition in heavy-ion collisions might occur around 4 GeV \cite{yonggcplb2024,yong_MethodProbing_2023a}. Therefore, for this study, we have set the energy range between 3 and 7.7 GeV. In this study, within the potential phase transition region of \(\sqrt{s_{NN}}=3-7.7\) GeV, we present the NCQ scaling behavior of hadron elliptic flow in heavy-ion collisions. Our research reveals that pure hadronic matter does not exhibit NCQ scaling behavior, while quark matter demonstrates a breakdown of NCQ scaling at beam energies below 3.9 GeV. Consequently, at lower energies, the NCQ scaling of hadron elliptic flow is no longer a reliable observable for determining the presence of quark matter.

%
The Multi-Phase Transport (AMPT) model, developed by Lin Ziwei et al., provides a comprehensive framework based on non-equilibrium many-body dynamics to study and interpret the results of relativistic heavy-ion collision experiments at RHIC and LHC energies~\cite{lin_MultiphaseTransport_2005}. Over the past two decades, the AMPT model has been extensively developed and validated, establishing itself as a reliable tool for reproducing and understanding a wide range of experimental observables in intermediate- and high-energy heavy-ion collisions. In this study, we utilize both the updated string melting (SM) version and the improved pure hadron cascade (HC) version of the AMPT model. The latter excludes partonic degrees of freedom and focuses solely on hadronic interactions~\cite{yong_DoubleStrangeness_2021, yong_MethodProbing_2023a}.

In the AMPT-SM model, the initial conditions are generated using the HIJING model~\cite{gyulassy_HIJING10_1994}, which initializes participants as hard minijet partons and soft strings. These strings are subsequently fragmented into partons, and the scattering between these partons is described by Zhang's Parton Cascade (ZPC) model~\cite{zhang_ZPC101_1998b}. To study heavy-ion collisions at low energies, the present AMPT-SM model includes the effects of finite nuclear thickness \cite{yong_MethodProbing_2023a,yonggcplb2024}. The total parton elastic scattering cross section is given by
\begin{equation}
    \sigma \approx \frac{9\pi \alpha_s^2}{2\mu^2},
\end{equation}
where \( \alpha_s \) is the strong coupling constant, and \( \mu \) is the screening mass. This cross section can be adjusted by modifying the screening mass or coupling constant. The hadronization process employs quark coalescence, wherein quark pairs (for mesons) or triplets (for baryons) combine when their relative distances and momenta in phase space simultaneously fall below predefined spatial and momentum coalescence radii. A simpler proximity criterion is adopted, utilizing coordinate-space thresholds of \(R_0 = 0.877 (0.61)\) fm for baryons (mesons) alongside momentum-space thresholds of \(P_0 = 0.89 (1.28)\) GeV/\(c\) for baryons (mesons). These parameters, derived from constituent quark model calculations and calibrated to approximate experimental proton and \(\pi^+\) radii, ensure phenomenological consistency. Among all eligible candidates, the quark configuration with the smallest phase-space separation is selected to form a hadron. The coalescence mechanism explicitly includes both ground and excited hadronic states such as \(\rho\), \(\omega\), \(K^*\), \(\phi\), \(\eta\), and \(\Delta\). During formation, three-momentum conservation is strictly enforced by setting the composite hadron's momentum equal to the vector sum of its constituent parton momenta. The invariant mass is then calculated via energy conservation, and the specific hadron species is determined by matching the parton flavor composition and computed invariant mass to known hadronic states. To ensure physical fidelity, the mass of the composite hadron is enforced to match the established experimental value for the identified species; any resultant mass discrepancy is compensated by adjusting the hadron's kinetic energy while rigorously preserving total energy conservation. The interactions between hadrons are then described using a hadronic cascade process based on an extended relativistic transport (ART) model~\cite{li_FormationSuperdense_1995b,lin_MultiphaseTransport_2005,yong_DoubleStrangeness_2021}.

In the AMPT-HC model, the initial positions and momenta of nucleons in the projectile and target nuclei are initialized using the Woods-Saxon density profile combined with the local Thomas-Fermi approximation. These nucleons directly undergo hadronic interactions, including both elastic and inelastic collisions, modeled using the same extended ART framework and hadron potentials with the test particle method \cite{yong_DoubleStrangeness_2021}. The key distinction between the AMPT-SM and AMPT-HC models lies in the inclusion or exclusion of partonic dynamics.

%
Figure~\ref{1} illustrates the NCQ scaling behavior of elliptic flow in the energy range of $\sqrt{s_{NN}} = 3.0$ to $7.7\,\mathrm{GeV}$ under two different AMPT modes. The transverse mass range shown in the figure is the same as that used by the STAR Collaboration in their analysis of hadron elliptic flow and NCQ scaling behavior at this beam energy region \cite{abdallah_DisappearancePartonic_2022}. In the SM mode, the scaling behavior becomes increasingly pronounced as the collision energy rises. At $\sqrt{s_{NN}} = 3.0\,\mathrm{GeV}$, the scaling is significantly violated, but it is largely restored when the energy exceeds $\sqrt{s_{NN}} = 3.9\,\mathrm{GeV}$. Although the SM mode incorporates a quark coalescence mechanism, the expected NCQ scaling is not observed at $\sqrt{s_{NN}} = 3.0\,\mathrm{GeV}$, which is somewhat unexpected. This suggests that NCQ scaling at several GeV may not necessarily serve as a definitive signature of partonic degrees of freedom. We further explore the reasons for the absence of quark scaling at $\sqrt{s_{NN}} = 3.0\,\mathrm{GeV}$ in the subsequent discussion.

In the HC mode, the elliptic flow as a function of transverse momentum ($p_T$) becomes smoother with increasing energy. However, a clear NCQ scaling behavior does not emerge, which aligns with the expectations of the quark coalescence mechanism. This is because NCQ scaling is generally interpreted as a consequence of the quark coalescence process. Since such a process is absent in the HC mode, the lack of NCQ scaling is understandable.

\begin{figure}[t]
\centering
\includegraphics[width=0.48\textwidth]{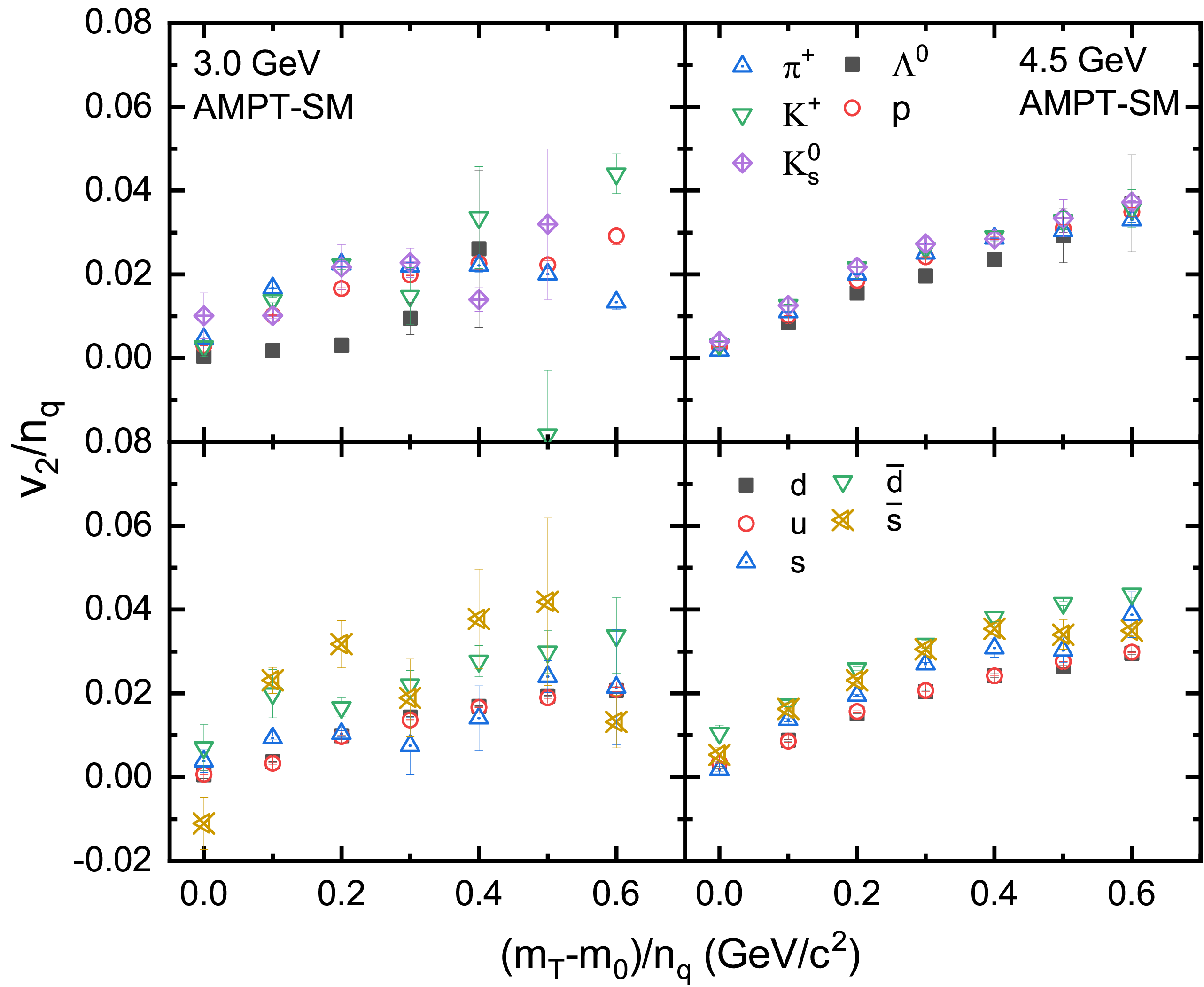}
\caption{The $v_{2}/n_{q}$ of different particles as a function of $(m_{T}-m_{0})/n_{q}$ at $\sqrt{s_{NN}}=3.0 \,\mathrm{GeV}$ and $\sqrt{s_{NN}}=4.5 \,\mathrm{GeV}$ in Au+Au collisions with a centrality of 10-40\%. Top: results for hadrons. Bottom: results for the constituent quarks of the hadrons shown in the top panels.} \label{2}
\vspace{0.25cm}
\end{figure}

\begin{figure}[t]
\centering
\vspace{-0.15cm}
\includegraphics[width=0.48\textwidth]{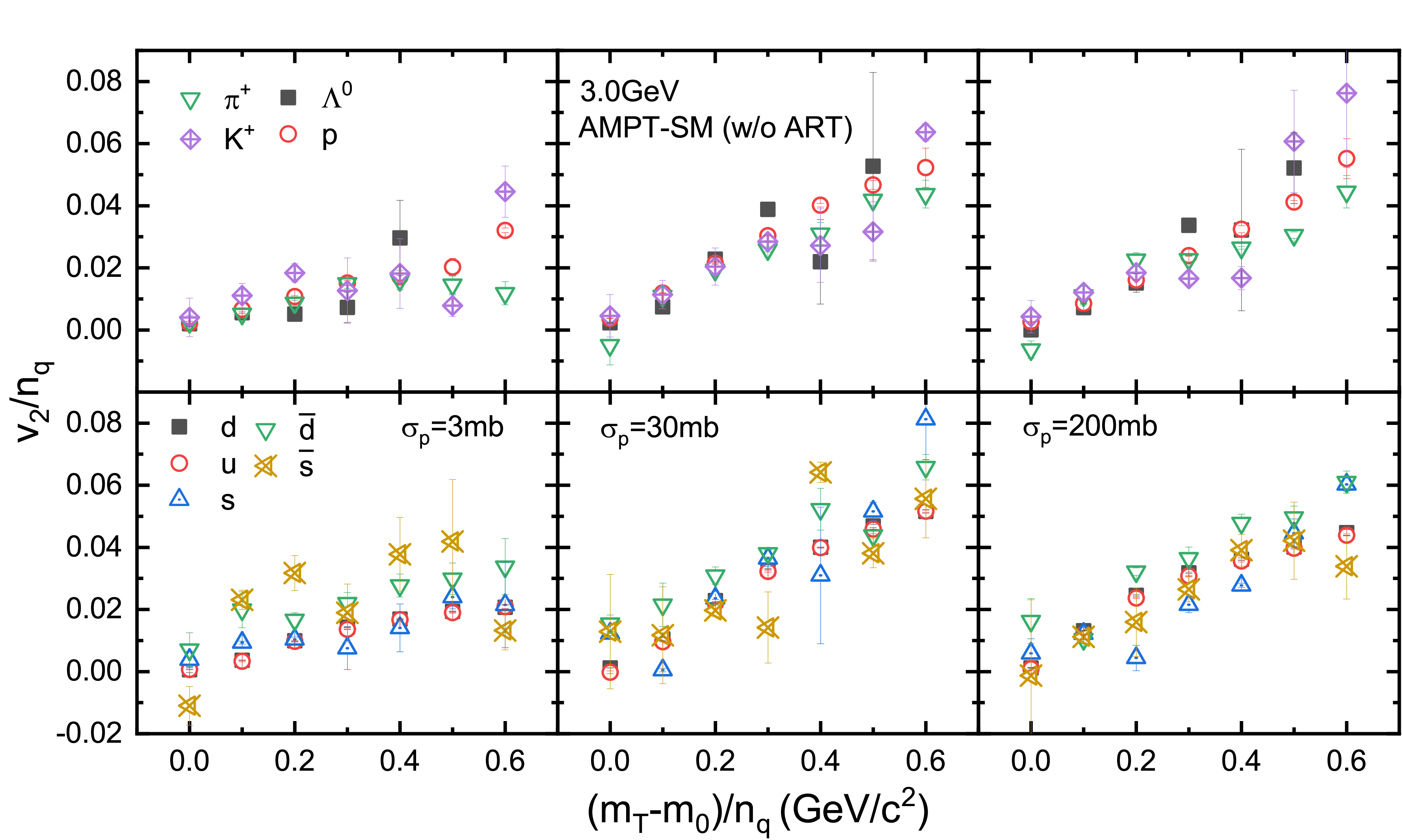}
\caption{The $v_{2}/n_{q}$ of different particles as a function of $(m_{T}-m_{0})/n_{q}$ in the same Au+Au collisions as shown in Figure~\ref{2}, but without the ART afterburner, at $\sqrt{s_{NN}}=3.0 \,\mathrm{GeV}$ with different parton scattering cross sections. Top: results for hadrons. Bottom: results for the constituent quarks of the hadrons shown in the top panels.} \label{3}
\vspace{0.25cm}
\end{figure}

Investigating the breakdown of NCQ scaling at $\sqrt{s_{NN}} = 3.0\,\mathrm{GeV}$ in the SM mode is crucial for understanding the quark coalescence mechanism and the origin of NCQ scaling. According to the quark coalescence model, the elliptic flow of hadrons arises from the elliptic flow of their constituent quarks. Figure~\ref{2} shows the elliptic flow of hadrons scaled by the number of constituent quarks, alongside the elliptic flow of quarks at $\sqrt{s_{NN}} = 3.0\,\mathrm{GeV}$ and $4.5\,\mathrm{GeV}$. At $\sqrt{s_{NN}} = 3.0\,\mathrm{GeV}$, the quark elliptic flow exhibits poor regularity, while at $\sqrt{s_{NN}} = 4.5\,\mathrm{GeV}$, it displays a more consistent scaling pattern. The regularity of the quark elliptic flow is transferred to the elliptic flow of hadrons through the quark coalescence process, thereby influencing the emergence of NCQ scaling at different energies. Since elliptic flow is sensitive to the degree of thermalization of the system~\cite{collaboration_EllipticFlow_2001, kolb_AnisotropicTransverse_2000e, voloshin_PhysicsCentrality_2000c}, we speculate that the breakdown of NCQ scaling in the quark elliptic flow at $\sqrt{s_{NN}} = 3.0\,\mathrm{GeV}$, as shown in the bottom panel of Figure~\ref{2}, may be related to insufficient thermalization. This hypothesis is supported by the results in Figure~\ref{3}, which depict the elliptic flow of quarks and hadrons under different parton scattering cross sections. Increasing the parton scattering cross section enhances the probability of parton collisions, thereby promoting system thermalization. To eliminate the influence of hadron scatterings on the analysis, we excluded hadron scatterings in Figure~\ref{3}. It is evident that as the parton scattering cross section increases, the quark elliptic flow exhibits improved scaling behavior, and the NCQ scaling of hadronic flow becomes more pronounced.

\begin{table}[t!]
    \centering
    \footnotesize
    \renewcommand{\arraystretch}{1.2}
    \label{quark}
    \resizebox{8cm}{!}{%
    \begin{tabular}{ccccccc}
        \hline\hline
        Energy & $d$ & $u$ & $s$ & $\bar{d}$ & $\bar{u}$ & $\bar{s}$ \\
        \hline
        3.0\,GeV & 118.0751 & 104.6948 & 0.175026 & 0.556889 & 0.458784 & 0.10041 \\
        4.5\,GeV & 125.6711 & 112.8621 & 7.231412 & 17.8743 & 17.25401 & 3.885175 \\
        \hline\hline
    \end{tabular}%
    }
    \caption{The number of quarks produced in the midrapidity region ($|y| \le 0.5$) in Au+Au collisions.}
    \label{quark}
\end{table}

\begin{figure}[t]
\centering
\includegraphics[width=0.48\textwidth]{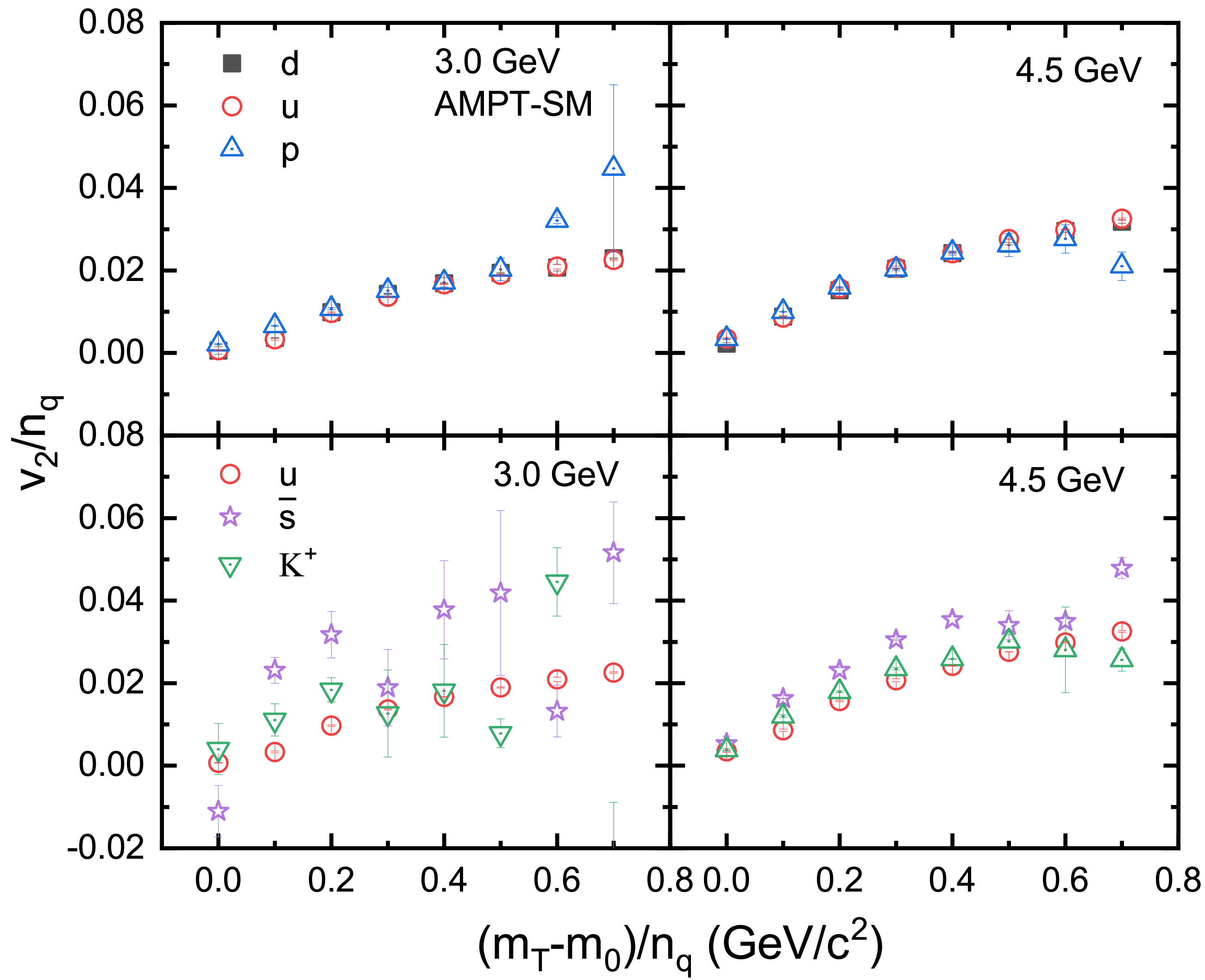}
\caption{Top: The $v_{2}/n_{q}$ of protons and their constituent quarks ($u$ and $d$) as a function of $(m_{T}-m_{0})/n_{q}$ in Au+Au collisions at $\sqrt{s_{NN}}=3.0 \,\mathrm{GeV}$ and $\sqrt{s_{NN}}=4.5 \,\mathrm{GeV}$. Bottom: The $v_{2}/n_{q}$ of $K^+$ mesons and their constituent quarks ($u$ and $\bar{s}$) as a function of $(m_{T}-m_{0})/n_{q}$ at $\sqrt{s_{NN}}=3.0 \,\mathrm{GeV}$ and $\sqrt{s_{NN}}=4.5 \,\mathrm{GeV}$.} \label{4}
\vspace{0.25cm}
\end{figure}

In addition to thermalization effects, our analysis reveals another critical factor. As shown in Table~\ref{quark}, the production of strange quarks and antiquarks at $\sqrt{s_{NN}} = 3.0\,\mathrm{GeV}$ is significantly lower by more than a factor of 30 compared to that at $\sqrt{s_{NN}} = 4.5\,\mathrm{GeV}$. The impact of this reduced strange quark abundance on NCQ scaling behavior is illustrated in Figure~\ref{4}. The upper panels of Figure~\ref{4} depict the elliptic flow of protons and their constituent quarks ($u$ and $d$). At both $\sqrt{s_{NN}} = 3.0\,\mathrm{GeV}$ and $4.5\,\mathrm{GeV}$, the yields of $u$ and $d$ quarks are substantial, and protons and their constituent quarks exhibit consistent scaling behavior. In contrast, the lower panels show the elliptic flow of $K^+$ mesons and their constituent quarks ($u$ and $\bar{s}$). At $\sqrt{s_{NN}} = 3.0\,\mathrm{GeV}$, the differential elliptic flow of strange quarks displays poor regularity compared to that at $\sqrt{s_{NN}} = 4.5\,\mathrm{GeV}$. This irregularity is likely due to the insufficient number of strange quarks at $\sqrt{s_{NN}} = 3.0\,\mathrm{GeV}$, which amplifies statistical fluctuations in their elliptic flow, resulting in irregular jumps. This irregularity is subsequently transmitted to the elliptic flow of $K^+$ mesons, causing it to also become irregular and preventing the emergence of NCQ scaling at $\sqrt{s_{NN}} = 3.0\,\mathrm{GeV}$. In contrast, at $\sqrt{s_{NN}} = 4.5\,\mathrm{GeV}$, the larger abundance of strange quarks supports a more robust collective flow, enabling the formation of NCQ scaling for $K^+$ mesons.
To sum up, within the energy range of $\sqrt{s_{NN}} = 3.0$--$7.7\,\mathrm{GeV}$, the emergence of NCQ scaling is influenced by multiple complex factors. Both the degree of quark thermalization and the overall quark abundance can affect the collective behavior of the quark system and, consequently, the manifestation of NCQ scaling in final-state hadrons. Therefore, in this energy region, NCQ scaling may no longer serve as a reliable indicator of QGP formation.

\begin{figure}[t]
\centering
\includegraphics[width=0.48\textwidth]{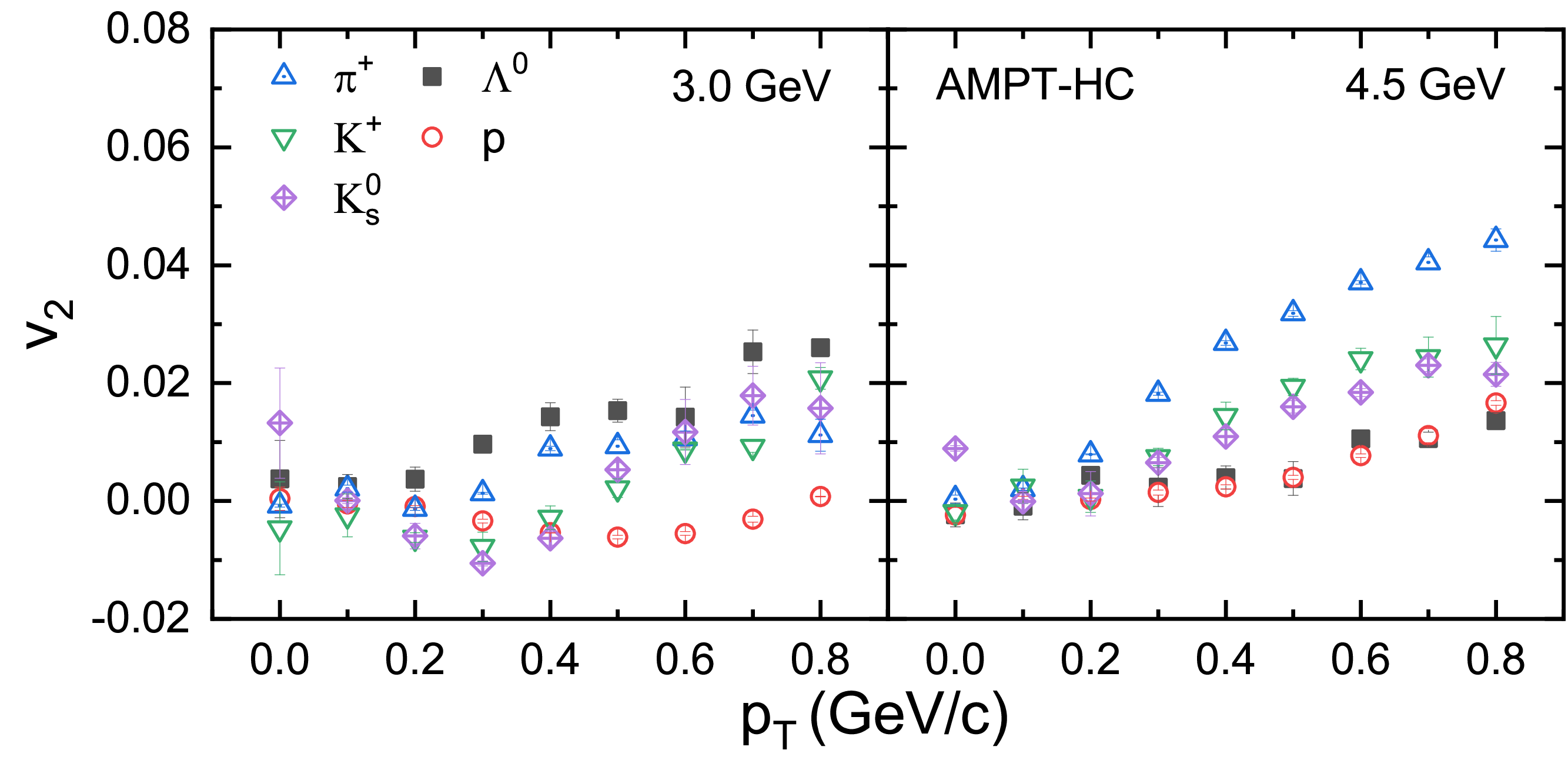}
\caption{The differential elliptic flow of different particles in Au+Au collisions with centrality 10-40\% at $\sqrt{s_{NN}}=3.0 \,\mathrm{GeV}$ and $\sqrt{s_{NN}}=4.5 \,\mathrm{GeV}$ in the AMPT-HC mode.} \label{5}
\vspace{0.25cm}
\end{figure}

\begin{figure}[t]
\centering
\includegraphics[width=0.48\textwidth]{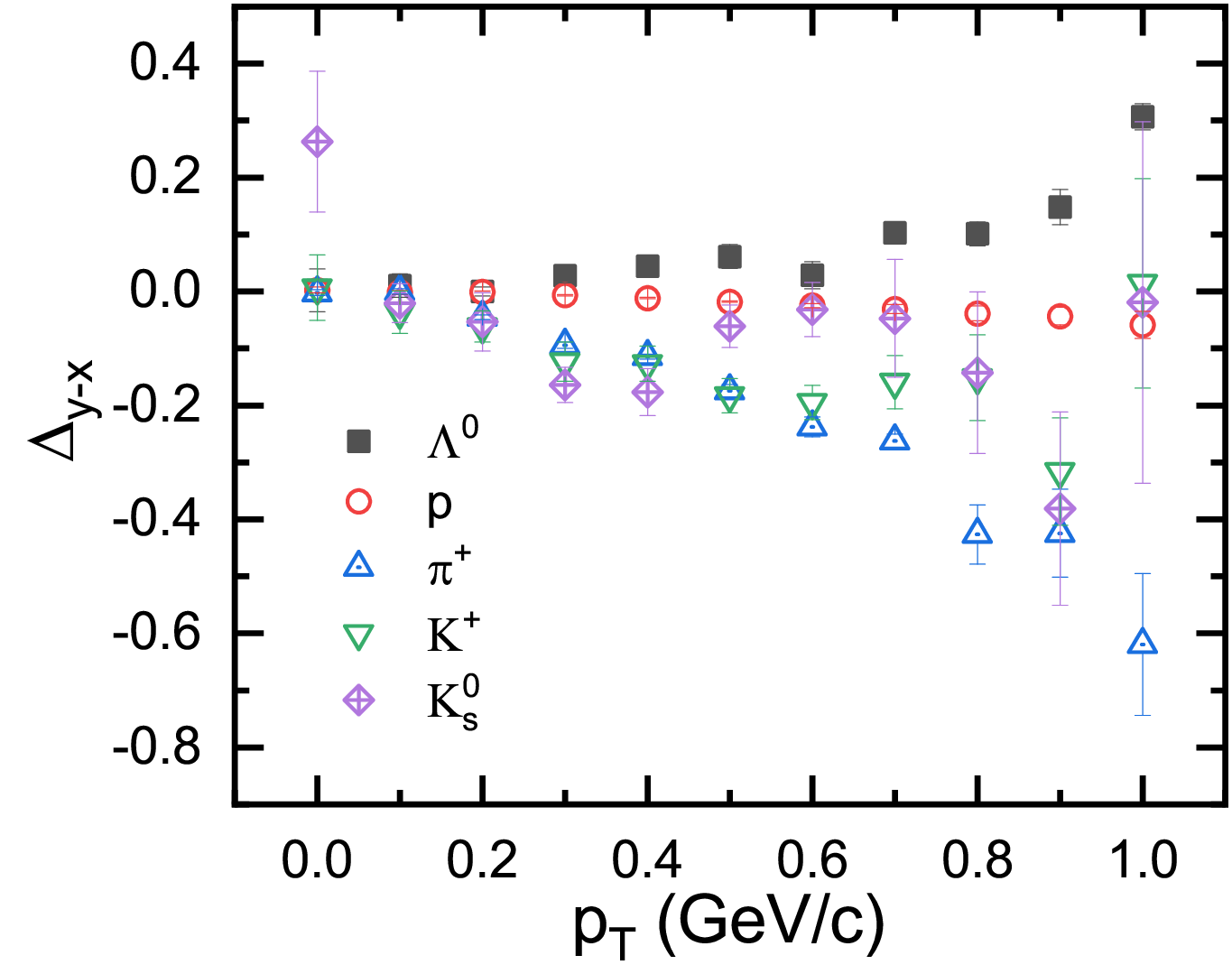}
\caption{The difference in particle number increment in Au+Au collisions at $\sqrt{s_{NN}}=4.5 \,\mathrm{GeV}$ compared to $\sqrt{s_{NN}}=3.0 \,\mathrm{GeV}$ in the $y$ and $x$ directions as a function of transverse momentum.} \label{6}
\vspace{0.25cm}
\end{figure}

To further analyze the results under the HC mode, we plot the differential elliptic flow at $\sqrt{s_{NN}} = 3.0\,\mathrm{GeV}$ and $4.5\,\mathrm{GeV}$, as shown in Figure~\ref{5}. There are two noteworthy features in Figure~\ref{5}. Firstly, an inversion in the elliptic flow of $\Lambda^{0}$ is observed between the two energies. This behavior may be related to both the properties of $\Lambda^{0}$ and the collective dynamics of the collision system. At $\sqrt{s_{NN}} = 3.0\,\mathrm{GeV}$, the beam energy is just above the $\Lambda^{0}$ production threshold, thus $\Lambda^{0}$ particles are produced in nucleon-nucleon collisions with relatively high transverse momenta. Due to the low collision energy and the small scattering cross sections of $\Lambda^{0}$ collides with other particles, the $\Lambda^{0}$ experiences few subsequent collisions, which allows their initially large momentum anisotropy to be largely preserved. As a result, the elliptic flow of $\Lambda^{0}$ is relatively high in this case. At $\sqrt{s_{NN}} = 4.5\,\mathrm{GeV}$, the higher collision energy leads to more frequent interactions, resulting in a more thermalized system with stronger collective behavior. Under such conditions, a clear mass ordering of elliptic flow emerges, in which heavier particles like $\Lambda^{0}$ exhibit smaller elliptic flow than lighter hadrons, leading to the inversion of the elliptic flow of $\Lambda^{0}$ between $\sqrt{s_{NN}} = 3.0\,\mathrm{GeV}$ and $\sqrt{s_{NN}} = 4.5\,\mathrm{GeV}$. Secondly, a clear mass ordering in the differential elliptic flow is observed at $\sqrt{s_{NN}} = 4.5\,\mathrm{GeV}$, while this behavior is absent at $\sqrt{s_{NN}} = 3.0\,\mathrm{GeV}$. This phenomenon has been well-documented in Au+Au collisions at $\sqrt{s_{NN}} = 130\,\mathrm{GeV}$ and $200\,\mathrm{GeV}$~\cite{adler_AzimuthalAnisotropy_2002a, collaboration_IdentifiedParticle_2001a, adams_ParticleTypeDependence_2004a} and has been successfully reproduced by hydrodynamic models~\cite{huovinen_RadialElliptic_2001a, borghini_MomentumSpectra_a, ollitrault_RelativisticHydrodynamics_2008}. In these models, mass ordering arises from the radial flow generated by the system's expansion, which imparts a collective boost to the momentum spectra of the produced particles. The magnitude of this effect depends on both the particle mass and the strength of the radial expansion. The quantitative impact of radial flow on $v_2$ is explicitly discussed in Eq.~(3) of Ref.~\cite{huovinen_RadialElliptic_2001a} and Eq.~(8) of Ref.~\cite{borghini_MomentumSpectra_a}, both of which confirm that elliptic flow decreases with increasing particle mass. Consequently, mass ordering is widely regarded as a signature of hydrodynamic expansion in heavy-ion collisions.

Interestingly, the mass ordering pattern is also reproduced in transport model simulations, as shown in the right panel of Figure~\ref{5}. This suggests that frequent scatterings among particles can generate radial flow, even in the absence of full thermalization. In this context, the emergence of mass ordering does not necessarily indicate that the system has reached the hydrodynamic limit. Since the strength of radial flow is correlated with the initial energy density~\cite{li_OriginMass_2017a}, the absence of mass ordering at $\sqrt{s_{NN}} = 3.0\,\mathrm{GeV}$ may be attributed to a lower initial energy density and, consequently, weaker collective expansion at this energy. Figure~\ref{6} illustrates the pushing effect of radial flow on particles of different masses. It shows the difference in the particle number increment with beam energy in the $y$-direction (perpendicular to the reaction plane) and the $x$-direction (along the reaction plane), defined as:
\begin{equation}
\Delta_{y-x} = \frac{N_y(4.5) - N_y(3.0)}{N_y(3.0)}-\frac{N_x(4.5) - N_x(3.0)}{N_x(3.0)}.
\end{equation}

As illustrated in Figure~\ref{6}, the difference between the y-direction and x-direction values increases with particle mass. This indicates that the impact of radial expansion varies depending on the particle type. Specifically, radial expansion enhances the likelihood of heavy particles appearing in the y-direction compared to lighter particles. At the same transverse momentum, the elliptic flow is proportional to the difference in particle yields along the y- and x-directions. When more particles are emitted in-plane (along the x-direction) than out-of-plane (along the y-direction), the elliptic flow is positive. Conversely, if the in-plane yield is suppressed, the elliptic flow value decreases. Consequently, heavier particles tend to exhibit smaller elliptic flow values at $\sqrt{s_{NN}} = 4.5\,\mathrm{GeV}$.

%
In summary, we employed both the string melting (SM) and hadron cascade (HC) versions of the AMPT model to investigate the number of constituent quark (NCQ) scaling of elliptic flow for \( p \), \( \Lambda^{0} \), \( \pi^{+} \), \( K^{+} \), and \( K^{0}_{S} \) in non-central Au+Au collisions at \(\sqrt{s_{NN}} = 3.0\text{--}7.7\,\mathrm{GeV}\). In the SM mode, NCQ scaling gradually emerges as the collision energy increases. At \(\sqrt{s_{NN}} = 3.0\,\mathrm{GeV}\), NCQ scaling is entirely absent, while at \(\sqrt{s_{NN}} = 4.5\,\mathrm{GeV}\), it is largely established. Although quark coalescence occurs at \(\sqrt{s_{NN}} = 3.0\,\mathrm{GeV}\), the absence of NCQ scaling can be attributed to two primary factors. First, the elliptic flow of quarks at this energy is not sufficiently developed, as shown in Figure~3. Increasing the parton scattering cross section to \(30\,\mathrm{mb}\) enhances the development of quark elliptic flow, making the NCQ scaling behavior of hadron elliptic flow more pronounced. Second, the limited number of strange quarks and antiquarks at \(\sqrt{s_{NN}} = 3.0\,\mathrm{GeV}\) results in weak collective effects, preventing NCQ scaling from manifesting in strange hadrons and mesons. These findings suggest that the presence or absence of NCQ scaling in elliptic flow cannot serve as a definitive indicator of quark-gluon plasma (QGP) formation at energies close to \(\sqrt{s_{NN}} = 3.0\,\mathrm{GeV}\). In the HC mode, no NCQ scaling is observed, as expected, since the quark coalescence mechanism is absent. However, a mass ordering phenomenon emerges in the elliptic flow at approximately \(\sqrt{s_{NN}} = 4.5\,\mathrm{GeV}\), indicating that mass ordering may not require full thermalization. In contrast, at \(\sqrt{s_{NN}} = 3.0\,\mathrm{GeV}\), no clear mass ordering is observed, likely due to the weaker collective effects at this energy.

In the investigation of the hadron-quark phase transition in heavy-ion collisions, researchers focus on qualitative observables such as NCQ scaling and net-proton number fluctuations because conventional observables in heavy-ion collisions are often susceptible to model uncertainties (this is also why, after nearly 30 years of research, the hadron-quark phase transition remains unconfirmed). Therefore, identifying observables for the hadron-quark phase transition that are less model-dependent --- such as the yield ratio of identical strange particles in dual-reaction systems, as we recently proposed as a signal for the hadron-quark phase transition \cite{zhux2025} --- remains an urgent task.

%
This work is supported by the National Natural Science Foundation of China under Grant Nos. 12275322, 12335008 and CAS Project for Young Scientists in Basic Research YSBR-088.

\end{document}